\documentclass[aps,prd,twocolumns,eqsecnum,preprintnumbers,showpacs,amsmath,amssymb]
{revtex4}

\usepackage{graphicx}
\usepackage{dcolumn}
\usepackage{bm}

\begin{document}

\title{THE EXISTENCE OF A MASS GAP IN QUANTUM YANG-MILLS THEORY}

\author{V. Gogohia}
\email[]{gogohia@rmki.kfki.hu}

\affiliation{HAS, CRIP, RMKI, Depart. Theor. Phys., Budapest 114,
P.O.B. 49, H-1525, Hungary}

\date{February 28, 2006}
\begin{abstract}

The skeleton loop integrals which contribute into the gluon
self-energy have been iterated (skeleton loops expansion) within
the Schwinger-Dyson equation for the full gluon propagator. No any
truncations/approximations as well as no special gauge choice have
been made. It is explicitly shown that such obtained general
iteration solution for the full gluon propagator can be exactly
and uniquely decomposed as a sum of the two principally different
terms. The first term is the Laurent expansion in integer powers
of severe (i.e., more singular than $1/ q^2$) infrared
singularities accompanied by the corresponding powers of the mass
gap and multiplied by the corresponding residues. The second
(perturbative) term is always as much singular as $1/q^2$ and
otherwise remaining undetermined. We have explicitly demonstrated
that the mass gap is hidden in the above-mentioned skeleton loop
integrals due to the nonlinear interaction of massless gluon
modes. It shows explicitly up when the gluon momentum goes to
zero. The appropriate regularization scheme has been applied in
order to make a gauge-invariant existence of the mass gap
perfectly clear. Moreover, it survives an infinite series
summation of the relevant skeleton loop contributions into the
gluon self-energy. The physical meaning of the mass gap is to be
responsible for the large scale structure of the true QCD vacuum.
\end{abstract}



\maketitle

\section{Introduction}

The Lagrangian of QCD \cite{1,2} does not contain explicitly any
of the mass scale parameters which could have a physical meaning
even after the corresponding renormalization program is performed.
However, if QCD itself is a confining theory then a characteristic
scale has to exist. It should be directly responsible for the
large scale structure of the true QCD vacuum in the same way as
$\Lambda_{QCD}$ is responsible for the nontrivial perturbative
dynamics there (scale violation, asymptotic freedom (AF)
\cite{1,2}). On one hand, the color confinement problem is not yet
solved \cite{3}. On the other hand, today there is no doubt that
color confinement is closely related to the above-mentioned
structure of the true QCD ground state \cite{4,5} and vice-versa.
The perturbation theory (PT) technics fail to investigate them.

 The main goal of this paper is to show how
the above-mentioned characteristic scale (the mass gap in what
follows, for simplicity) responsible for the nonperturbative (NP)
dynamics in the infrared (IR) region may explicitly appear in QCD.
This especially becomes imperative after Jaffe and Witten have
formulated their theorem "Yang-Mills Existence And Mass Gap"
\cite{6}. Moreover, we will show that the mass gap may not only
appear, but it may also survive after an infinite series summation
of the relevant skeleton loop contributions into the gluon
self-energy. Thus the paper is devoted to a possible solution of
one of the important problems in theoretical particle/nuclear
physics, namely to the dynamical generation of a mass gap in
quantum field gauge theories.

The propagation of gluons is one of the main dynamical effects in
the true QCD vacuum. The gluon Green's function is (Euclidean
signature here and everywhere below)

\begin{equation}
D_{\mu\nu}(q) = i \left\{ T_{\mu\nu}(q)d(q^2, \xi) + \xi
L_{\mu\nu}(q) \right\} {1 \over q^2 },
\end{equation}
where $\xi$  is the gauge-fixing parameter and
$T_{\mu\nu}(q)=\delta_{\mu\nu}-q_{\mu} q_{\nu} / q^2 =
\delta_{\mu\nu } - L_{\mu\nu}(q)$. Evidently, $T_{\mu\nu}(q)$ is
the transverse ("physical") component of the full gluon
propagator, while $L_{\mu\nu}(q)$ is its longitudinal (unphysical)
one. The free gluon propagator is obtained by setting simply the
full gluon form factor to $d(q^2, \xi)=1$ in Eq. (1.1), i.e.,

\begin{equation}
D^0_{\mu\nu}(q) = i \left\{ T_{\mu\nu}(q) + \xi L_{\mu\nu}(q)
\right\} {1 \over q^2}.
\end{equation}

The main tool of our investigation is the Schwinger-Dyson (SD)
equation of motion \cite{1,7,8} (and references therein) for the
full gluon propagator (1.1), since its solutions reflect the
quantum-dynamical structure of the true QCD ground state. Some
results of the present investigation have been already briefly
announced in our preliminary publications \cite{9,10}. Here we
present the full (without any estimates,
approximations/truncations, specific gauge choice, etc.)
investigation of the problem of the dynamical origin of a mass gap
in quantum Yang-Mills (YM) theory.

\section{Gluon SD equation}

The general structure of the gluon SD equation \cite{1,7,9} can be
written down symbolically as follows (for our purposes it is more
convenient to consider the SD equation for the full gluon
propagator and not for its inverse):

\begin{equation}
D(q) = D^0(q) + D^0(q)T_g[D](q)D(q).
\end{equation}
Here and in some places below, we omit the dependence on the Dirac
indices, for simplicity, as well as the quark and ghost skeleton
loop contributions into the gluon self-energy (as it is required
by the YM character of our consideration).  The nonlinear (NL)
pure gluon contribution $T_g[D](q)$ into the gluon self-energy is
the sum of the four topologically independent skeleton loop
integrals, namely

\begin{equation}
T_g[D](q)  = {1 \over 2} T_t + {1 \over 2} T_1(q) + {1 \over 2}
T_2(q) + {1 \over 6} T_2'(q),
\end{equation}
where the so-called constant tadpole term is

\begin{equation}
T_t =  g^2 \int {i d^4 q_1 \over(2 \pi)^4} T^0_4 D(q_1),
\end{equation}
and all other skeleton loop integrals are given explicitly below
as follows:

\begin{equation}
T_1(q) =  g^2 \int {i d^4 q_1 \over (2 \pi)^4} T^0_3 (q, -q_1,
q_1-q) T_3 (-q, q_1, q -q_1) D(q_1) D(q -q_1),
\end{equation}

\begin{equation}
T_2(q) =  g^4 \int {i d^4 q_1 \over (2 \pi)^4} \int {i d^n q_2
\over (2 \pi)^4} T^0_4 T_3 (-q_2, q_3, q_2 -q_3) T_3(-q, q_1,
q_3-q_2) D(q_1) D(-q_2)D(q_3) D(q_3 -q_2),
\end{equation}

\begin{equation}
T_2'(q) =  g^4 \int {i d^4 q_1 \over (2 \pi)^4} \int {i d^4 q_2
\over (2 \pi)^4} T^0_4 T_4 (-q, q_1, -q_2, q_3) D(q_1)
D(-q_2)D(q_3),
\end{equation}
where in the last two skeleton loop integrals $q-q_1 +q_2-q_3=0$
as usual. Evidently, neither the color group factors nor the Dirac
indices play any role in tracking down the mass gap, which can
only be of dynamical origin. That is the reason why they are not
explicitly shown throughout this paper.

 The general iteration solution (i.e., when the
above-mentioned skeleton loop integrals are to be iterated) of the
gluon SD equation (2.1) looks like

\begin{eqnarray}
D(q)&=& D^{(0)}(q) + D^{(1)}(q)+ D^{(2)}(q) + D^{(3)}(q) + ...
\nonumber\\
&=& D^0(q) + D^0(q)T_g[D^0+D^{(1)}+ D^{(2)} + D^{(3)} + ... ](q)
[D^0(q)+
D^{(1)}(q) + D^{(2)}(q)+ D^{(3)}(q) + ...], \nonumber\\
\end{eqnarray}
and $D^{(0)}(q) = D^0(q)$. It is nothing but the skeleton loops
expansion, and it is not the PT series. First of all, the
magnitude of the coupling constant squared cannot be fixed to be
small, i.e., it is arbitrary. Secondly, the dependence of the
skeleton loop integrals on the coupling constant squared is also
completely arbitrary, i.e., it cannot be explicitly fixed on
general ground. However, this expansion is rather formal, since
the corresponding skeleton loop integrals are not yet regularized.
In the deep IR limit $q^2 \rightarrow 0$ the skeleton loop
integrals (2.4)-(2.6) tend to their corresponding divergent
constant values (in Euclidean metrics $q^2 \rightarrow 0$ implies
$q_i \rightarrow 0$). Just these constants having the dimensions
of a mass squared are the main objects of our investigation. The
only problem is as how to extract them, i.e., to make their
existence and important role perfectly clear. Fortunately, this
can be explicitly done through the necessary regularization of the
initial skeleton loop integrals.

\section{Regularization}

Due to AF \cite{1,2} all the skeleton loop integrals as well as
those which will appear in the formal iteration solution (2.7) are
divergent, so the general problem of their regularization arises.
For our future purpose it is convenient to regularize the
above-mentioned skeleton loop integrals by subtracting as usual
their values at a safe (slightly different from zero) space-like
point $q^2 = \mu^2$ (Euclidean signature is already chosen). Thus,
one obtains

\begin{eqnarray}
T_1^R(q) &=& T_1(q) - T_1(q^2 = \mu^2), \nonumber\\
T_2^R(q) &=& T_2(q) - T_2(q^2 = \mu^2), \nonumber\\
T_2'^R(q) &=& T_2'(q) - T_2'(q^2 = \mu^2),
\end{eqnarray}
where all the divergent constants \underline{having the dimensions
of a mass squared} are

\begin{eqnarray}
T_1(q^2 = \mu^2) = g^2 \int {i d^4 q_1 \over (2 \pi)^4} T^0_3 (q,
-q_1, q_1-q) T_3 (-q, q_1, q -q_1) D(q_1) D(q -q_1)
\arrowvert_{q^2= \mu^2}, \nonumber\\
T_2(q^2 = \mu^2) =  g^4 \int {i d^4 q_1 \over (2 \pi)^4} \int {i
d^n q_2 \over (2 \pi)^4} T^0_4 T_3 (-q_2, q_3, q_2 -q_3) T_3(-q,
q_1, q_3-q_2) \nonumber\\
D(q_1) D(-q_2)D(q_3) D(q_3 -q_2) \arrowvert_{q^2 = \mu^2},
\nonumber\\
T_2'(q^2 = \mu^2) = g^4 \int {i d^4 q_1 \over (2 \pi)^4} \int {i
d^4 q_2 \over (2 \pi)^4} T^0_4 T_4 (-q, q_1, -q_2, q_3) D(q_1)
D(-q_2)D(q_3) \arrowvert_{q^2 = \mu^2}.
\end{eqnarray}
At the same time, the introduction of the ultraviolet (UV) cutoff
$\Lambda^2$ in these integrals is assumed . All this makes it
possible first of all to explicitly release the above-mentioned
divergent constants, and put all the divergent skeleton loop
integrals under firm control. The UV cutoff should go to infinity
at the final stage only. The formal subtraction procedure for the
constant tadpole term (2.3) implies $T^R_t = T_t - T_t =0$. This
is in agreement with the dimensional regularization method
\cite{1,2} where $T_t(D^0)=0$, indeed. So, we can discard this
term within the formal iteration solution (2.7) without loosing
generality.

The subtractions in Eq. (3.1) mean that the decomposition of the
regularized quantities into the independent tensor structures can
be written down as follows:

\begin{eqnarray}
T^R_1(q) \equiv T^R_{(1)\mu\nu}(q) = \delta_{\mu\nu} q^2
T_1^{(1)}(q^2) + q_{\mu} q_{\nu} T_1^{(2)}(q^2), \nonumber\\
T^R_2(q) \equiv T^R_{(2)\mu\nu}(q) = \delta_{\mu\nu} q^2
T_2^{(1)}(q^2) + q_{\mu} q_{\nu} T_2^{(2)}(q^2), \nonumber\\
T_2'^R(q) \equiv T_{(2)\mu\nu}'^R(q) = \delta_{\mu\nu} q^2
T_2'^{(1)}(q^2) + q_{\mu} q_{\nu} T_2'^{(2)}(q^2).
\end{eqnarray}
where all invariant functions $T_1^{(n)}(q^2), T_2^{(n)}(q^2),
T_2'^{(n)}(q^2) $ at $n=1,2$ are dimensionless ones. In the region
of small $q^2$ they are represented in the form of the
corresponding Taylor expansions and remain arbitrary otherwise.
Due to the definition $q_{\mu} q_{\nu} =q^2 L_{\mu\nu}$, instead
of the independent structures $\delta_{\mu\nu}$ and $q_{\mu}
q_{\nu}$ in Eqs. (3.3) and below, one can use $T_{\mu\nu}$ and
$L_{\mu\nu}$ as the independent structures with their own
invariant functions. From these relations it follows that the
regularized skeleton loop contributions are always of the order
$q^2$, i.e., $T^R_1(q) = O(q^2), \ T^R_2(q) = O(q^2)$ and
$T_2'^R(q) = O(q^2)$.

The NL pure gluon part (2.2) thus can be exactly decomposed as the
sum of the two terms:

\begin{equation}
T_g[D](q)  = T_g[D] + T^R_g[D](q),
\end{equation}
where

\begin{equation}
T_g[D]  =  {1 \over 2} T_1(D^2) + {1 \over 2} T_2(D^4) + {1 \over
6} T_2'(D^3),
\end{equation}
and

\begin{equation}
T^R_g[D](q)  = {1 \over 2} T^R_1(q) + {1 \over 2} T^R_2(q) + {1
\over 6} T_2'^R(q) = O(q^2; D).
\end{equation}
In Eq. (3.5) we introduce the following notations: $T_1(q^2 =
\mu^2)= T_1(D^2), \ T_2(q^2 = \mu^2)= T_2(D^4)$ and $T_2'(q^2 =
\mu^2)= T_2'(D^3)$, showing their dependence on the corresponding
number of the gluon propagators only. In Eq. (3.6) the last
equality shows explicitly that its left-hand-side is always of the
order $q^2$, depending again on $D$ in general. Thus the gluon SD
equation (2.1) becomes divided into the two principally different
terms, namely

\begin{equation}
D(q) = D^0(q) + D^0(q)T_g[D]D(q) + D^0(q)O(q^2;D)D(q).
\end{equation}
Let us note that both blocks, $T_g[D]$ and $O(q^2;D)$ (which
should be iterated with respect to $D$), depend on $D$ themselves.
This means that there is no way to sum up in general these
iteration series into the corresponding geometric progression. The
latter possibility appears only when the blocks to be iterated do
not depend on $D$ like in quantum electrodynamics (QED), where the
electron skeleton loop is only present. The difference between
these two cases lies, of course, in the NL dynamics of QCD in
comparison with the linear one in QED.

Let us emphasize once more that the constant block $T_g[D]$ having
the dimensions of a mass squared is just the object we have
worried about to demonstrate explicitly its crucial role within
our approach. In this connection a few additional remarks are in
order. As it follows from the standard gluon SD equation (3.7),
the corresponding equation for the gluon self-energy looks like

\begin{equation}
D^{-1}(q) = D^{-1}_0(q) - T_g[D] - O(q^2;D),
\end{equation}
where we put $D^0(q) \equiv D_0(q)$. In order to unravel
overlapping UV divergence problems in YM theory, the necessary
number of the differentiation with respect to the external
momentum should be done first (in order to lower divergences).
Then the point-like vertices, which are present in the skeleton
loop integrals, should be replaced by their full counterparts via
the corresponding integral equations. Finally, one obtains the
corresponding SD equations which are much more complicated that
the standard ones, containing different scattering amplitudes,
which skeleton expansions are, however, free from the
above-mentioned overlapping divergences. Of course, the real
procedure \cite{11} (and references therein) is much more tedious
than briefly described above. However, even at this level, it is
clear that by taking derivatives with respect to the external
momentum $q$ in Eq. (3.8), the main initial information due to the
constant block $T_g[D]$ will be totally lost. Whether it will be
restored somehow or not at the later stages of the renormalization
program is not clear at all. Thus in order to remove overlapping
UV divergences ("the water") from the SD equations and skeleton
expansions, we are in danger to completely loose the information
on the constant block $T_g[D]$ which is the dynamical source of
the mass gap ("the baby") within our approach. In order to avoid
this danger and to be guaranteed that no any dynamical information
are lost, we are just using the standard gluon SD equation (3.7).
The presence of any kind of UV divergences (overlapping and usual
(overall)) in the skeleton expansions will not cause any problems
in order to detect the mass gap responsible for the IR structure
of the true QCD vacuum. In other words, the direct iteration
solution of the standard gluon SD equation (3.7), complemented by
the proposed regularization scheme, is reliable to release a mass
gap, and thus to make its existence perfectly clear. The problem
of convergence of such regularized skeleton series is completely
irrelevant in the context of the present investigation. Anyway, we
keep any kind of UV divergences under control within our method.
At the same time, any kind of UV divergences play no any role in
the existence of a mass gap responsible for the IR structure of
the full gluon propagator, i.e., its existence does not depend on
whether overlapping divergences are present or not in the SD
equations and corresponding skeleton expansions. All this is the
main reason why our starting point is the standard gluon SD
equation (2.1) for the unrenormalized Green's functions (this also
simplifies notations).

\section{Nonlinear iteration}

The formal iteration solution of the gluon SD equation (3.7) now
looks like

\begin{eqnarray}
D(q)&=& D^{(0)}(q) + D^{(1)}(q)+ D^{(2)}(q) + D^{(3)}(q) + ... =
D^0(q) \nonumber\\
&+& D^0(q)T_g[D^0+D^{(1)}+ D^{(2)} + D^{(3)}(q) + ... ][D^0(q)+
D^{(1)}(q) + D^{(2)}(q) + D^{(3)}(q) + ...] \nonumber\\
&+& D^0(q)O(q^2; D^0+D^{(1)}+ D^{(2)} + D^{(3)}(q) + ...
)[D^0(q)+ D^{(1)}(q) + D^{(2)}(q) + D^{(3)}(q) + ...], \nonumber\\
\end{eqnarray}
where

\begin{eqnarray}
D^{(0)}(q) &=& D^0(q), \nonumber\\
D^{(1)}(q) &=& D^0(q) F_1[D^0]D^0(q) +  D^0(q)O(q^2;D^0)D^0(q), \nonumber\\
D^{(2)}(q) &=& D^0(q) F_2[D^0, D^{(1)}][D^0(q) + D^{(1)}(q)] +
D^0(q) O(q^2; D^0 + D^{(1)})[D^0(q) + D^{(1)}(q)],
\nonumber\\
D^{(3)}(q) &=& D^0(q) F_3[D^0, D^{(1)}, D^{(2)}][D^0(q) +
D^{(1)}(q) + D^{(2)}(q)] \nonumber\\
&+& D^0(q) O(q^2; D^0 + D^{(1)} + D^{(2)}) [D^0(q) + D^{(1)}(q) +
D^{(2)}(q)],
\end{eqnarray}
and so on. Let us consider the divergent constants $F_1[D^0],
F_2[D^0, D^{(1)}], F_3[D^0, D^{(1)}, D^{(2)}],...$, introducing
the following notations:

\begin{eqnarray}
F_1[D^0] &=& T_g[D^0]  = {1 \over 2} T_1((D^0)^2) + {1 \over 2}
T_2((D^0)^4) + {1 \over 6} T_2'((D^0)^3), \nonumber\\
F_2[D^0, D^{(1)}] &=& T_g[D^0+D^{(1)}] = {1 \over 2}
T_1((D^0+D^{(1)})^2) + {1 \over 2} T_2((D^0+D^{(1)})^4) + {1 \over
6} T_2'((D^0+D^{(1)})^3),
\nonumber\\
F_3[D^0, D^{(1)}, D^{(2)}] &=& T_g[D^0+D^{(1)}+D^{(2)}] = {1 \over
2} T_1((D^0+D^{(1)}+D^{(2)})^2) \nonumber\\
&+& {1 \over 2} T_2((D^0+D^{(1)}+D^{(2)})^4) + {1 \over 6}
T_2'((D^0+D^{(1)}+D^{(2)})^3),
\end{eqnarray}
and so on. As underlined above, each of them has the dimensions of
a mass squared, so on general ground one can represent them as
follows:

\begin{eqnarray}
F_1 \equiv F_1[D^0] &=& \Delta^2 C_1(\lambda, \nu, \xi, g^2),
\nonumber\\
F_2 \equiv F_2[D^0, D^{(1)}] &=& \Delta^2 C_2(\lambda, \nu,
\xi, g^2), \nonumber\\
F_3 \equiv F_3[D^0, D^{(1)}, D^{(2)}] &=& \Delta^2 C_3(\lambda,
\nu, \xi, g^2),
\end{eqnarray}
and so on. In these relations $\Delta^2$ is the above-mentioned
mass gap. The dimensionless constants $C_1,C_2,C_3, ...$ depend on
the dimensionless UV cutoff $\lambda$ introduced as follows:
$\Lambda^2 = \lambda \Delta^2$. They depend on the dimensionless
renormalization point $\nu$ introduced as follows: $\mu^2 = \nu
\Delta^2$. Evidently, via the corresponding subscripts these
constants depend on which iteration for the gluon propagator $D$
is actually done, which in its turn gives the dependence on the
gauge-fixing parameter $\xi$. They also depend on the
dimensionless coupling constant squared $g^2$ (see Eqs. (3.2)).
The parameters $\lambda$ and $\nu$ may depend on $\xi$ and $g^2$
(and hence vise-versa). The dependence of the mass gap $\Delta^2$
on all these parameters is not shown explicitly, for convenience,
but can be restored any time, if necessary.

In the relations (4.3) and (4.4) we use the short-hand notation
$D^0 \equiv D^0(q)$, however, it is more appropriate to introduce
the short-hand notations as follows:

\begin{eqnarray}
D^0(q) &\equiv& D_0(q) \equiv D_0, \ O_1(q^2) \equiv O(q^2;D^0), \nonumber\\
O_2(q^2) &\equiv& O(q^2; D^0 + D^{(1)}), \ O_3(q^2) \equiv O(q^2;
D^0 + D^{(1)} + D^{(2)}),
\end{eqnarray}
and so on. Then the formal iteration solution (4.1) becomes

\begin{eqnarray}
D(q)&=& D^{(0)}(q) + D^{(1)}(q)+ D^{(2)}(q) + D^{(3)}(q) + ... =
D_0 + [D_0 F_1D_0 +  D_0 O_1(q^2)D_0] \nonumber\\
&+& [D_0 F_2D_0 + D_0 F_2D_0F_1D_0 + D_0 F_2D_0 O_1(q^2)D_0
\nonumber\\
&+& D_0 O_2(q^2)D_0 +D_0 O_2(q^2)D_0F_1D_0 + D_0
O_2(q^2)D_0O_1(q^2)D_0] \nonumber\\
&+& [D_0F_3D_0 + D_0F_3D_0F_1D_0 + D_0F_3D_0 O_1(q^2)D_0
\nonumber\\
&+& D_0F_3D_0F_2D_0 + D_0F_3D_0F_2D_0F_1D_0 +
D_0F_3D_0F_2D_0O_1(q^2)D_0 \nonumber\\
&+&D_0F_3D_0O_2(q^2)D_0 + D_0F_3D_0O_2(q^2)D_0F_1D_0
+D_0F_3D_0O_2(q^2)D_0O_1(q^2)D_0 \nonumber\\
&+& D_0O_3(q^2)D_0 + D_0O_3(q^2)D_0F_1D_0 +
D_0O_3(q^2)D_0O_1(q^2)D_0 \nonumber\\
&+&D_0O_3(q^2)D_0F_2D_0 + D_0O_3(q^2)D_0F_2D_0F_1D_0 +
D_0O_3(q^2)D_0F_2D_0O_1(q^2)D_0 \nonumber\\
&+&D_0O_3(q^2)D_0O_2(q^2)D_0+D_0O_3(q^2)D_0O_2(q^2)D_0F_1D_0 +
D_0O_3(q^2)D_0O_2(q^2)D_0O_1(q^2)D_0] + ... \ . \nonumber\\
\end{eqnarray}

\section{Shifting procedure}

The formal iteration series (4.6), however, much more convenient
to equivalently rewrite as follows:

\begin{eqnarray}
D(q)&=& [D_0^2 (F_1+F_2+F_3+ ...) +  D_0^3(F_1F_2 +F_3F_1+ F_2F_3+
...) + D_0^4(F_1F_2F_3 + ...) + ... ] \nonumber\\
&+& [D_0 F_2D_0 O_1(q^2)D_0 + D_0 O_2(q^2)D_0F_1D_0 + D_0F_3D_0
O_1(q^2)D_0 + D_0F_3D_0F_2D_0O_1(q^2)D_0 \nonumber\\
&+& D_0F_3D_0O_2(q^2)D_0 + D_0F_3D_0O_2(q^2)D_0F_1D_0
+D_0F_3D_0O_2(q^2)D_0O_1(q^2)D_0 \nonumber\\
&+& D_0O_3(q^2)D_0F_1D_0 + D_0O_3(q^2)D_0F_2D_0 +
D_0O_3(q^2)D_0F_2D_0F_1D_0 \nonumber\\
&+& D_0O_3(q^2)D_0F_2D_0O_1(q^2)D_0 +
D_0O_3(q^2)D_0O_2(q^2)D_0F_1D_0 + ... ] \nonumber\\
&+& [D_0 + D_0 O_1(q^2)D_0 + D_0 O_2(q^2)D_0 + D_0 O_3(q^2)D_0 +
D_0O_2(q^2)D_0O_1(q^2)D_0 \nonumber\\
&+&D_0O_3(q^2)D_0O_1(q^2)D_0+D_0O_3(q^2)D_0O_2(q^2)D_0 +
D_0O_3(q^2)D_0O_2(q^2)D_0O_1(q^2)D_0 + ...]. \nonumber\\
\end{eqnarray}

Since $D_0 \equiv D_0(q) \sim (q^2)^{-1}$, there are three
formally different types of terms. The first type of terms is
singular as much as $D_0^{2+k} \sim (q^2)^{-2-k}, \
k=0,1,2,3,...$. Evidently, only the divergent constants
$F_1,F_2,F_3,...$ and their combinations enter into these terms.
The third type of terms is singular as much as the free gluon
propagator, i.e., they are of the order $D_0 \sim (q^2)^{-1}$,
since all functions $O_n(q^2), n=1,2,3,...$ are of the order
$q^2$. Evidently, only the functions $O_n(q^2), n=1,2,3,...$ and
their products enter into these terms. The second type of terms is
the so-called mixed up terms which contain the divergent constants
$F_1,F_2,F_3,...$ and the functions $O_n(q^2), n=1,2,3,...$ in the
different combinations.

Let us show that any mixed up term is simply the exact sum of the
first and third types of terms. In order to show this explicitly,
let us recall that all functions $O_n(q^2), n=1,2,3,...$ are
regular functions of $q^2$. So, we can equivalently represent them
in the form of the corresponding Taylor expansions. It is
convenient to present such kind of expansions in terms of
$D_0^{-1} \sim q^2$ and making an exact decomposition as follows:

\begin{eqnarray}
O_n(q^2) &=& O_n(q^2) - D_0^{-1} \sum_{m=0}^{k-1} (D_0
\Delta^2)^{-m} O_n^{(m)}(1) + D_0^{-1} \sum_{m=0}^{k-1} (D_0
\Delta^2)^{-m}O_n^{(m)}(1) \nonumber\\
&=& D_0^{-1} \sum_{m=0}^{k-1} (D_0 \Delta^2)^{-m} O_n^{(m)}(1) +
D_0^{-1}(D_0 \Delta^2)^{-k} f_n(q^2),  \quad n=1,2,3,... \ .
\end{eqnarray}
Here $f_n(q^2)$ are dimensionless functions with constant behavior
at zero momentum and otherwise remaining arbitrary. The last term
always has the corresponding order in powers of $D_0^{-1}$, so
that

\begin{eqnarray}
D_0^{-1}(D_0 \Delta^2)^{-k} f_n(q^2)  &=& O_n(q^2) - D_0^{-1}
\sum_{m=0}^{k-1} (D_0
\Delta^2)^{-m} O_n^{(m)}(1) \nonumber\\
&=& D_0^{-1} \sum_{m=0}^{\infty} (D_0 \Delta^2)^{-m} O_n^{(m)}(1)
- D_0^{-1} \sum_{m=0}^{k-1} (D_0 \Delta^2)^{-m} O_n^{(m)}(1) \sim
D_0^{-1}(D_0 \Delta^2)^{-k}, \nonumber\\
\end{eqnarray}
indeed. The same is true for the products of two, three, and more
different $O_n(q^2)$ functions (see below).

For the first mixed up term, $D_0F_2D_0O_1(q^2)D_0$, one then gets

\begin{eqnarray}
D_0 F_2D_0 O_1(q^2)D_0 &=& D_0^{2+1}F_2 [D_0^{-1} O_1^{(0)}(1)
+D_0^{-1}(D_0 \Delta^2)^{-1} f_1(q^2) ] \nonumber\\
&=& D_0^2F_2 O_1^{(0)}(1) + D_0 F_2 \Delta^{-2} f_1(q^2) =
D_0^2F_2O_1^{(0)}(1) + O(D_0),
\end{eqnarray}
since for this term we should put $k=1$ in Eq. (5.2). The
combination of the mass squared parameters $F_2 \Delta^{-2}$ is
the dimensionless one (see relations (4.4)), and the last term is
obviously denoted as $O(D_0)$. Thus, one concludes that this mixed
up term becomes the exact sum of the two different terms. The
first one should be included into the terms of the order $D_0^2$
shown as the first term in the iteration solution (5.1), while the
second term should be combined with the terms which always are of
the order $D_0$.

For the mixed up term, $D_0F_3D_0F_2D_0O_1(q^2)D_0$, we obtain

\begin{eqnarray}
D_0F_3D_0 F_2D_0 O_1(q^2)D_0 &=& D_0^{2+2}F_2 F_3[D_0^{-1}
O_1^{(0)}(1) + D_0^{-2} \Delta^{-2}O_1^{(1)}(1) + D_0^{-1}(D_0
\Delta^2)^{-2} f_1(q^2)] \nonumber\\
&=& D_0^3F_2 F_3 O_1^{(0)}(1) + D_0^2 F_2F_3
\Delta^{-2}O_1^{(1)}(1) + D_0 F_3F_2 \Delta^{-4}f_1(q^2) \nonumber\\
&=& D_0^3 F_2F_3 O_1^{(0)}(1) + D_0^2 F_2 F_3 \Delta^{-2}
O_1^{(1)}(1) + O(D_0),
\end{eqnarray}
since for this term we should put $k=2$ in Eq. (5.2). The
combination $F_2 F_3 \Delta^{-2}$ has the dimensions of a mass
squared (see Eqs. (4.4)), while the combination $F_2 F_3
\Delta^{-4}$ is dimensionless. Again the last term is denoted as
$O(D^0)$. So this term also becomes the exact sum of the three
terms. The first term from this sum has to be shifted into the
second term of Eq. (5.1). The second term from this sum has to be
shifted into the first term of Eq. (5.1), and the last one has to
be shifted into the last term of the general iteration solution
(5.1).

It is instructive to consider in more details the terms which
contain two and more $O(q^2)$ functions.  The corresponding Taylor
expansions for the product of any two and three functions are as
follows:

\begin{eqnarray}
O_n(q^2)O_l(q^2) &=& D_0^{-2} \sum_{m=0}^{k-2} (D_0 \Delta^2)^{-m}
O_{nl}^{(m)}(1) + D_0^{-2}(D_0 \Delta^2)^{-k+1} f_{nl}(q^2), \nonumber\\
O_n(q^2)O_l(q^2) O_j(q^2) &=& D_0^{-3} \sum_{m=0}^{k-3} (D_0
\Delta^2)^{-m} O_{nlj}^{(m)}(1) + D_0^{-3}(D_0 \Delta^2)^{-k+2}
f_{nlj}(q^2),
\end{eqnarray}
and so on. Here $f_{nl}(q^2)$ and $f_{nlj}(q^2)$ are dimensionless
functions with constant behavior at small momentum $q^2$ and
otherwise remaining arbitrary. Similarly to the previous case, the
last terms are the terms of the corresponding orders in powers of
$D_0^{-1}$, so that

\begin{eqnarray}
D_0^{-2}(D_0 \Delta^2)^{-k+1} f_{nl}(q^2) &\sim&
O_{nl}(D_0^{-k-1}) \sim D_0^{-2}(D_0 \Delta^2)^{-k+1},
\nonumber\\
D_0^{-3}(D_0 \Delta^2)^{-k+2} f_{nlj}(q^2) &\sim&
O_{nlj}(D_0^{-k-1}) \sim D_0^{-3}(D_0 \Delta^2)^{-k+2},
\end{eqnarray}
indeed, and so on. Then, for example, for the mixed up term,
$D_0O_3(q^2)D_0 F_2D_0 O_1(q^2)D_0$, one gets

\begin{eqnarray}
D_0O_3(q^2)D_0 F_2D_0 O_1(q^2)D_0 &=& D_0^{2+2}F_2 [D_0^{-2}
O_{31}^{(0)}(1) + D_0^{-2}(D_0 \Delta^2)^{-1} f_{31}(q^2)  ] \nonumber\\
&=& D_0^2F_2 O_{31}^{(0)}(1) + D_0 F_2 \Delta^{-2} f_{31}(q^2) \nonumber\\
&=& D_0^2 F_2 O_{31}^{(0)}(1) + O(D_0),
\end{eqnarray}
since for this term we should put $k=2$ in the first of Eqs.
(5.6). The combination $F_2 \Delta^{-2}$ is dimensionless, and the
last term is denoted as $O(D_0)$. Again this mixed up term becomes
the exact sum of two terms. The first term should be shifted into
the first, while the second one should be shifted into the last
term of the general iteration solution (5.1).

Moreover, all other mixed up terms, which are explicitly present
and omitted in formal series (5.1), should be treated in the same
way. Let us underline that the exact and unique separation between
the two kind of terms ($\sim D_0^{2+k}(q), \ k=0,1,2,3,...$ and
$\sim D_0(q))$ is achieved by keeping the necessary number of
terms in the corresponding Taylor expansions. This "shifting"
method in its general form has been formulated and applied in our
previous publications \cite{7,10}. It is worth emphasizing that by
shifting we do not change the functional dependence (and hence the
dependence on the mass gap) in the terms $\sim D_0^{2+k}(q)$, only
the accompanied $q^2$-independent factors will be changed. At the
same time, the terms $\sim D_0(q))$ will be changed (many new
terms of the same order will appear). Let us also note that all
the combinations of different masses squared (like the
above-mentioned $F_2 \Delta^{-2}, F_2F_3 \Delta^{-2}$, etc.) can
be reduced to the corresponding powers of the mass gap $\Delta^2$
and dimensionless coefficients $C_1, C_2, C_3,...$ via the
relations (4.4). They are to be multiplied by the different
dimensionless constants $O_n^{(m)}(1), \ O_{nl}^{(m)}(1), \
O_{nlj}^{(m)}(1)$, etc., which appear through the shifting
procedure, and which in general may depend on the same quantities
as in Eq. (4.4).

Rearranging all the terms, one gets that the general iteration
solution (5.1) for full gluon propagator becomes the exact sum of
the two different terms, namely

\begin{equation}
D(q)= D_0^2(q) \Delta^2 \sum_{k=0}^{\infty}[D_0(q) \Delta^2]^k
\sum_{m=0}^{\infty} a_{k,m}(\lambda, \nu, \xi, g^2) + O(D_0(q)).
\end{equation}
It is worth emphasizing once more that Eq. (5.9) obtained by the
shifting procedure is equivalent to the initial general iteration
series (5.1). In other words, the shifting method is not an
approximation, but the exact method of the corresponding
rearrangement of the terms. It makes it possible to represent
initial series (5.1) in the form much more convenient for our
purpose (to track down the mass gap), while shifting functional
ambiguity of the initial series (5.1) (which is due to the
functions $O_n(q^2)$) into the term $O(D_0(q))$ in Eq. (5.9). Let
us recall that the terms $O(D_0(q))$ is the sum of the terms which
are of the order $D_0(q)$ from the very beginning (the third type
of terms in the expansion (5.1)) and the terms of the same order
having appeared due to the above-described shifting procedure.

\section{Exact structure of the full gluon propagator}

Restoring the tensor structure, omitting the tedious algebra and
again taking into account that $D_0 (q) \sim (q^2)^{-1}$, the
general iteration solution of the gluon SD equation (5.9) for the
full gluon propagator can be algebraically (i.e., exactly)
decomposed as the sum of the two principally different terms as
follows:

\begin{eqnarray}
D_{\mu\nu}(q) &=& D^{INP}_{\mu\nu}(q, \Delta^2)+
D^{PT}_{\mu\nu}(q) \nonumber\\
&=& i T_{\mu\nu}(q) {\Delta^2 \over (q^2)^2} \sum_{k=0}^{\infty}
(\Delta^2 / q^2)^k \sum_{m=0}^{\infty} \phi_{k,m}(\lambda, \nu,
\xi, g^2) + i \Bigr[ T_{\mu\nu}(q) \sum_{m=0}^{\infty} a_m(q^2;
\xi) + \xi L_{\mu\nu}(q) \Bigl] {1 \over q^2},
\end{eqnarray}
where the superscript "INP" stands for the intrinsically NP part
of the full gluon propagator. Its exact structure inevitably stems
from the general iteration solution of the standard gluon SD
equation. The important feature of our method is that the skeleton
loop integrals have been iterated (skeleton loops expansion), so
no any assumptions and approximations have been made. We
distinguish between two terms in Eq. (6.1) by the character of the
corresponding IR singularities and the explicit presence of the
mass gap (see below). It is worth emphasizing that both terms are
valid in the whole energy/momentum range, i.e., they are not
asymptotics. At the same time, we achieved the exact separation
between two terms responsible for the NP (dominating in the IR
($q^2 \rightarrow 0$)) and PT (dominating in the UV ($q^2
\rightarrow \infty$)) dynamics in the true YM vacuum.

The PT part of the full gluon propagator remains undetermined. The
exact dependence of the PT gluon form factor $d^{PT}(q^2, \xi) =
\sum_{m=0}^{\infty} a_m(q^2, \xi)$ on $q^2$ cannot be fixed on
general ground like it has been done in its INP counterpart (what
we only know about $a_m(q^2, \xi)$-functions is that all of them
are regular at small $q^2$, and the sum over them produces the PT
logarithm improvements at large $q^2$ due to AF). Evidently, the
presence of overlapping and overall UV divergences in the PT part
of the full gluon propagator cannot change the structure of its
INP part, i.e., its functional dependence on $q^2$ and hence its
dependence on the mass gap. They may only affect the
$q^2$-independent factors $\phi_{k,m}(\lambda, \nu, \xi, g^2)$,
which concrete values, however, are not important. In the PT part
the sum over $m$ indicates that all iterations contribute into the
PT IR singularity only, which is defined as always being as much
singular as the power-type IR singularity of the free gluon
propagator $(q^2)^{-1}$. That is why the longitudinal component of
the full gluon propagator should be included into its PT part.
Anyway, we are not responsible for this part. It is the prize we
have payed to fix exactly the functional dependence of the INP
part of the full gluon propagator. In Refs. \cite{7,9,10} we came
to the same structure (6.1) but in a rather different ways.

\subsection{The INP phase in QCD}

The exact decomposition of the full gluon propagator into the two
principally different terms in Eq. (6.1) is only possible on the
basis of the corresponding decomposition of the full gluon form
factor $d(q^2, \xi)$ in Eq. (1.1), namely

\begin{equation}
d(q^2, \xi) = d(q^2, \xi) - d^{PT}(q^2, \xi) + d^{PT}(q^2, \xi) =
d^{NP}(q^2, \xi) + d^{PT}(q^2, \xi).
\end{equation}
Let us note that in principle the full gluon form factor can be
defined as the effective charge of QCD, i.e., $d(q^2) =
\alpha_s(q^2)$, where the dependence on $\xi$ is omitted, for
simplicity. This algebraic decomposition makes it possible to
define correctly the NP phase in comparison with the PT one. The
full gluon form factor $d(q^2, \xi)$ being the NP effective
charge, nevertheless, is "contaminated" by the PT contributions,
while $d^{NP}(q^2, \xi)$ is the truly NP one, since it is free of
them, by construction \cite{7,10}. Substituting the exact
decomposition (6.2) into the full gluon propagator (1.1), one
obtains

\begin{equation}
D_{\mu\nu}(q) = i \left\{ T_{\mu\nu}(q)d(q^2, \xi) + \xi
L_{\mu\nu}(q) \right\} {1 \over q^2 } = D^{INP}_{\mu\nu}(q)+
D^{PT}_{\mu\nu}(q),
\end{equation}
where

\begin{equation}
D^{INP}_{\mu\nu}(q) = i T_{\mu\nu}(q) d^{NP}(q^2, \xi) {1 \over
q^2} =i T_{\mu\nu}(q) d^{INP}(q^2, \xi),
\end{equation}
and

\begin{equation}
D^{PT}_{\mu\nu}(q) = i \Bigr[ T_{\mu\nu}(q) d^{PT}(q^2, \xi)+ \xi
L_{\mu\nu}(q) \Bigl] {1 \over q^2},
\end{equation}
in complete agreement with Eq. (6.1), indeed.

As it follows from Eq. (6.1), the INP part of the full gluon
propagator in Eq. (6.4) is nothing else, but the corresponding
Laurent expansion in integer powers of $q^2$ accompanied by the
corresponding powers of the mass gap squared and multiplied by the
sum over the $q^2$-independent factors, namely

\begin{equation}
D^{INP}_{\mu\nu}(q, \Delta^2) = i T_{\mu\nu}(q)
\sum_{k=0}^{\infty} (q^2)^{-2-k} (\Delta^2)^{1+k} \phi_k(\lambda,
\nu, \xi, g^2),
\end{equation}
where

\begin{equation}
\phi_k(\lambda, \nu, \xi, g^2) = \sum_{m=0}^{\infty}
\phi_{k,m}(\lambda, \nu, \xi, g^2)
\end{equation}
are the so-called residues at poles. The sum over $m$ indicates
that an infinite number of iterations (all iterations) of the
corresponding skeleton loop integrals invokes each severe (for
definition see below) IR singularity labelled by $k$.

\vspace{3mm}

Using the exact decomposition of the full gluon propagator
described above, we can define in general terms the INP phase in
QCD as follows:

\vspace{3mm}

{\bf (i).} \ The INP phase is characterized by the presence of the
power-type severe (or equivalently NP) IR singularities
$(q^2)^{-2-k}, \ k=0,1,2,3,...$. So these IR singularities are
defined as more singular than the power-type IR singularity of the
free gluon propagator $(q^2)^{-1}$. The Laurent expansion (6.6)
necessarily starts from the simplest NP IR singularity
$(q^2)^{-2}$ possible in four-dimensional QCD, indeed \cite{7}.

\vspace{3mm}

{\bf (ii).} \ It depends only on the transverse ("physical")
degrees of freedom of gauge bosons.

\vspace{3mm}

{\bf (iii).} \ It is gauge-invariant. Though the coefficients
$\phi_{k,m}(\lambda, \nu, \xi, g^2)$ of the Laurent expansion
(6.6) may explicitly depend on the gauge-fixing parameter $\xi$,
the structure of this expansion itself does not depend on it (see
discussion below as well).

\vspace{3mm}

{\bf (iv).} \ The INP part of the full gluon propagator vanishes
as the mass gap goes to zero, while the PT part survives.

\vspace{3mm}

Within our approach the mass gap determines the power-type
deviation of the full gluon propagator from the free one in the IR
limit ($q^2 \rightarrow 0$), while $\Lambda_{QCD}$ determines the
logarithmic deviation of the full gluon propagator from the free
one in the UV limit ($q^2 \rightarrow \infty$). So, we distinguish
between the two different phases in QCD not by the strength of the
coupling constant squared (which is arbitrary in our approach),
but rather by the explicit presence of the mass gap, in which case
the coupling constant plays no any role. The INP phase disappears
when it goes to zero even if the NP IR singularities are not
explicitly present. So the subtraction (6.2) can be equivalently
written down as follows:

\begin{equation}
d^{NP}(q^2, \Delta^2) = d(q^2, \Delta^2) - d(q^2, \Delta^2=0) =
d(q^2, \Delta^2) - d^{PT}(q^2),
\end{equation}
i.e., it goes to zero as $\Delta^2 \rightarrow 0$, indeed. This
once more emphasizes the important role of the mass gap in the
definition of the truly NP phase as a particular case of the INP
one when the NP IR singularities are not explicitly present as
stated above. In other words, the existence of the truly NP phase
in any approach based on the gluon propagator assumes the regular
dependence on the mass scale parameter, which is chosen to play
the role of the mass gap. Otherwise, in the absence of the mass
gap and in order to recover the truly NP phase the UV asymptotic
of the full gluon propagator should be subtracted in agreement
with the relation (6.8). It is worth emphasizing once more that
the existence of the mass gap and the presence of the NP IR
singularities in the Laurent expansion (6.6) (and hence in the
full gluon propagator (6.1)) is absolutely general phenomenon. It
does not depend on the concrete values of the parameters:
$\lambda, \ \nu, \ \xi, \ g^2$. It is only due to the NL
interaction of massless gluons.

\section{Discussion}

The unavoidable presence of the first term in Eq. (6.1) makes the
principal difference between non-abelian QCD and abelian QED,
where such kind of term in the full photon propagator is certainly
absent (in the former theory there is direct coupling between
massless gluons, while in the latter one there is no direct
coupling between massless photons). Precisely this term may violet
the cluster properties of the Wightman functions \cite{12}, and
thus validates the Strocchi theorem \cite{13}, which allows for
such IR singular behavior of the full gluon propagator.

The INP part of the full gluon propagator in the form of the
corresponding Laurent expansion describes the so-called zero
momentum modes enhancement (ZMME or simply ZME which means zero
momentum enhancement) effect in the true QCD vacuum due to the NL
dynamics of massless gluon modes there. As underlined above, we do
not specify explicitly the value of the gauge-fixing parameter
$\xi$. So the ZMME effect takes place at its any value. In this
sense this effect is gauge-invariant. This is very similar to AF.
It is well known that the exponent which determines the
logarithmic deviation of the full gluon propagator from the free
one in the UV region ($q^2 \gg \Lambda^2_{QCD}$) explicitly
depends on the gauge-fixing parameter. At the same time, AF itself
does not depend on it, i.e., it takes place at any $\xi$.

Also, due to the arbitrariness of the above-mentioned residues
$\phi_k(\lambda, \nu, \xi, g^2)$ there is no way to sum up these
Laurent series into the function with regular behavior at small
$q^2$. However, the smooth in the IR gluon propagator is also
possible depending on different truncations/approximations used,
since the gluon SD equation is highly nonlinear one. The number of
solutions for such kind of systems is not fixed $a \ priori$. The
singular and smooth in the IR solutions for the gluon propagator
are independent from each other, and thus should be considered on
equal footing. Anyway, in order to find the smooth gluon
propagator completely different (from direct iteration) method of
the solution of the gluon SD equation should be used \cite{8,14}
(and references therein as well), since we have explicitly shown
here in a gauge-invariant way and making no any
approximations/truncations that the general iteration solution is
inevitably severely singular at small gluon momentum.

The QCD Lagrangian does not contain a mass gap. However, we
discovered that the mass scale parameter responsible for the NP
dynamics in the IR region exists in the true QCD ground state. At
the level of the gluon SD equation it is hidden in the skeleton
loop contributions into the gluon self-energy. It explicitly shows
up (and hence the corresponding severe IR singularities) when the
gluon momentum goes to zero. At the fundamental quark-gluon (i.e.,
Lagrangian) level the dynamical source of the mass gap is the
triple and quartic gluon vertices, i.e., the NL dynamics of QCD.
The former vanishes when all the gluon momenta involved go to zero
($T_3(0,0) =0$), while the latter survives in the same limit
($T_4(0,0,0) \neq 0$). Because of these features it would be
tempting to think that the quartic potential $(A \wedge A)^2$ in
the action plays much more important role than its triple
counterpart in the arising of severe IR singularities in quantum
YM theory. However, since we are dealing with the skeleton loops
expansion, we are unable (at least at this stage) to exactly
establish that the mass gap could arise from the quartic potential
rather than from its triple counterpart \cite{6} (see also Ref.
\cite{15}). So this dilemma remains an open but an interesting
problem to be solved.

In this connection, it is necessary to discuss an interesting
feature of the INP part of the full gluon propagator. Its
functional dependence on $q^2$ and hence on the mass gap (i.e.,
the Laurent structure of the expansion in Eq. (6.6)) is exactly
established up to the corresponding residues $\phi_k(\lambda, \nu,
\xi, g^2)$. In these residues the contributions from both the
triple and quartic gluon vertices have been taken into account.
However, the Laurent structure of the INP part does not depend on
whether we will take into account only three- or only four-gluon
vertices or both of them in the above-mentioned skeleton loop
integrals. We cannot omit both, so one of the NL interactions
should be always present. On the other hand, the residues
themselves will depend, of course, on the character of the NL
interaction taken into account. However, obviously, their concrete
values are not important, only the general dependence of the
residues $\phi_k(\lambda, \nu, \xi, g^2)$ on their arguments is
all that matters (in the subsequent paper we will show this
explicitly). At the same time, the dependence of the functions
$a_m(q^2; \xi)$ and hence of the PT form factor $d^{PT}(q^2; \xi)$
in Eq. (6.1) on $q^2$ will heavily depend on the character of the
NL interactions taken into account in the corresponding skeleton
loop integrals but this is not important for us as underlined
above.

Thus the true QCD vacuum is really beset with severe IR
singularities. They should be summarized (accumulated) into the
full gluon propagator and effectively correctly described by its
structure in the deep IR domain, exactly represented by its INP
part. The second step is to assign a mathematical meaning to the
integrals, where such kind of severe IR singularities will
explicitly appear, i.e., to define them correctly in the IR region
\cite{7,9,10}. This can be done by the use of the dimensional
regularization method \cite{16} correctly implemented into the
distribution theory \cite{17} (see subsequent paper). Just this IR
violent behavior makes QCD as a whole an IR unstable theory, and
therefore it has no IR stable fixed point, indeed \cite{1,18}.
This means that QCD itself might be a confining theory without
involving some extra degrees of freedom \cite{1,18,19} (and
references therein).

There is no doubt that our solution for the full gluon propagator,
obtained at the expense of remaining unknown its PT part,
nevertheless, satisfies the gluon SD equation (3.7), since it has
been obtained by the direct iteration solution of this equation.
To show this explicitly by substituting it back into the initial
gluon SD equation is not a simple task, however, and this is to be
done elsewhere. The problem is that the decomposition of the full
gluon propagator into the INP and PT parts by regrouping the
so-called mixed up terms in section V was a well defined procedure
(there was an exact criterion introduced in section VI how to
distinguish between these two terms in a single $D$). However, to
do the same at the level of the gluon SD equation itself, which is
nonlinear in $D$, is not so obvious. Also, the corresponding
severe IR singularities should be put under control at first
within the distribution theory (all this will be explicitly
demonstrated in a forthcoming paper).

\section{Conclusions}

A few years ago Jaffe and Witten (JW) have formulated the
following theorem \cite{6}:

\vspace{3mm}

 {\bf Yang-Mills Existence And Mass Gap:} Prove that
for any compact simple gauge group $G$, quantum Yang-Mills theory
on $\bf{R}^4$ exists and has a mass gap $\Delta > 0$.

\vspace{3mm}

Of course, at present to prove the existence of the YM theory with
compact simple gauge group $G$ is a formidable task yet. It is
rather mathematical than physical problem. However, one of the
main results of our investigation here can be formulated similar
to the above-mentioned JW theorem as follows:

\vspace{5mm}

{\bf Mass Gap Existence:} If quantum Yang-Mills theory with
compact simple gauge group $G=SU(3)$ exists on $\bf{R}^4$, then it
exhibits a mass gap.

\vspace{3mm}

Our mass gap $\Delta^2$ remains neither IR nor UV renormalized
yet, since at this stage it has been only regularized, i.e.,
$\Delta^2 \equiv \Delta^2(\lambda, \nu, \xi, g^2)$. However, there
is no doubt that it will survive both renormalization programs
(see subsequent paper). So denoting its IR and UV renormalized
version in advance as $\Lambda_{NP}$, then a symbolic relation
between it, the JW mass gap ($\Delta \equiv \Delta_{JW}$) and
$\Lambda_{QCD} \equiv \Lambda_{PT}$ could be written as

\begin{equation}
\Lambda_{NP} \longleftarrow^{\infty \leftarrow \alpha_s}_{0
\leftarrow M_{IR}} \ \Delta_{JW} \ { }^{\alpha_s \rightarrow
0}_{M_{UV} \rightarrow \infty} \longrightarrow  \ \Lambda_{PT}.
\end{equation}
Here $\alpha_s$ is obviously the fine structure coupling constant
of strong interactions, while $M_{UV}$ and $M_{IR}$ are the UV and
IR cut-offs, respectively. The right-hand-side limit is well known
as the weak coupling regime, while the left-hand-side can be
regarded as the strong coupling regime. We know how to take the
former \cite{1,2,18}, and we hope that we have explained here how
to deal with the latter one, not solving the gluon SD equation
directly, which is a formidable task, anyway. However, there is no
doubt that the final goal of this limit, namely, the mass gap
$\Lambda_{NP}$ exists, and should be renormalization group
invariant in the same way as $\Lambda_{QCD}$. It is solely
responsible for the large scale structure of the true QCD ground
state, while $\Lambda_{PT}$ is responsible for the nontrivial PT
dynamics there.

It is important to emphasize once more that the mass gap has not
been introduced by hand. We have explicitly demonstrated that it
is hidden in the skeleton loop integrals contributing into the
gluon self-energy due to the NL interaction of massless gluon
modes (Eqs. (2.4)-(2.6)). The mass gap shows explicitly up when
the gluon momentum goes to zero. An appropriate regularization
scheme has been applied to make the existence of the mass gap
perfectly clear. Moreover, it survives an infinite series
summation of the corresponding skeleton loop contributions
(skeleton loop expansion). In other words, an infinite number of
iterations of the relevant skeleton loops has to be made in order
to exhibit a mass gap. No any truncations/approximations have been
made as well as no special gauge choice, i.e., the result of the
summation is gauge-invariant. In the presence of the mass gap the
QCD coupling constant plays no any role. All its orders contribute
into the mass gap (skeleton loop expansion). This explains why the
interaction in our picture can be considered as a strong one. It
is worth emphasizing that our mass gap and the JW mass gap cannot
be interpreted as the gluon mass, i.e., they always remain
massless. These features point out on a possible similarity
between our mass gap and the JW one. Moreover, in a next paper we
will explicitly show (by investigating the IR renormalization
properties of the mass gap) that the interaction in our picture is
not only strong but short-ranged as well. This will allow us to
analytically formulate the gluon confinement criterion in a
gauge-invariant way.

Our second important result, a byproduct of the proof of the
existence of a mass gap, is that the general iteration solution of
the gluon SD equation is unavoidably severely IR singular at small
gluon momentum. The exactly established Laurent structure of the
INP part of the full gluon propagator (6.1) clearly shows that an
infinite number of iterations of the relevant skeleton loops
(skeleton loops expansion) invokes each NP IR singularity. Again
no special gauge choice and no any truncations/approximations have
been made as well in such obtained general iteration solution of
the gluon SD equation for the full gluon propagator. So, our gluon
propagator (more precisely its INP part) takes into account the
importance of the quantum excitations of severely singular IR
degrees of freedom in the true QCD vacuum. They lead to the
formation of the purely transverse quantum virtual field
configurations with the enhanced low-frequency components/large
scale amplitudes due to the NL dynamics of the massless gluon
modes. We will call them the purely transverse singular gluon
fields, for simplicity. In the presence of such severe IR
singularities the IR multiplicative renormalization (IRMR) program
is needed to perform in order to remove them in a self-consistent
way from the theory (see subsequent paper). In another forthcoming
paper we will show that the quark and ghost degrees of freedom
play no any significant role in the dynamical generation of a mass
gap. The NL interaction of massless gluon modes is only important
within our approach.

In summary, the existence of a mass gap in quantum YM theory has
been proved in a gauge-invariant way. We have explicitly shown
(again in a gauge-invariant way) that the direct iteration
solution of the gluon SD equation is inevitably severely singular
at small gluon momentum. Our general conclusion is that the
behavior of QCD at large distances is governed by the mass gap,
and therefore it should play a crucial role in the NL realization
of the quantum-dynamical mechanism of color confinement.

\begin{acknowledgments}

Support in part by HAS-JINR and Hungarian OTKA-T043455 grants is
to be acknowledged. I would like to thank P. Forgacs for bringing
my attention to the JW presentation of the Millennium Prize
Problem in Ref. \cite{6}. It is a great pleasure to thank A. Jaffe
for bringing my attention to the revised version of the
above-mentioned presentation. The author is also grateful to D.V.
Shirkov, A.N. Tavkhelidze, G.V. Efimov, D. Kazakov, J. Nyiri, Gy.
Pocsik, P. Levai, T. Biro, P. Vecsernyes and especially V.A.
Rubakov for useful discussions, remarks, suggestions and help at
the preliminary stages of this investigation.

\end{acknowledgments}

\end{document}